\documentclass[sigconf]{acmart}
\renewcommand\footnotetextcopyrightpermission[1]{} 
\usepackage[ruled,linesnumbered]{algorithm2e} 
\usepackage{subfig} 
\usepackage{multirow}
\usepackage{booktabs} 
\usepackage{threeparttable} 

\usepackage{tcolorbox}
\usepackage{float}
\usepackage{pifont}
\usepackage{ulem}
\usepackage{color, colortbl}

\begin{document}

\title{When LLM Meets Hypergraph: A Sociological Analysis on Personality via Online Social Networks }


\author{Zhiyao Shu}
\authornote{Equal contribution}
\affiliation{
  \institution{UC Berkeley, USA}
  \country{}
}
\email{yaoshu0326@berkeley.edu}

\author{Xiangguo Sun}
\authornotemark[1]
\authornote{Corresponding author}
\affiliation{%
  \institution{The Chinese University of Hong Kong}
  \country{}
  }
\email{xiangguosun@cuhk.edu.hk}

\author{Hong Cheng}
\affiliation{
  \institution{The Chinese University of Hong Kong}
  \country{}
  }
\email{hcheng@se.cuhk.edu.hk}

\begin{abstract}
Individual personalities significantly influence our perceptions, decisions, and social interactions, which is particularly crucial for gaining insights into human behavior patterns in online social network analysis. 
Many psychological studies have observed that personalities are strongly reflected in their social behaviors and social environments. \textbf{Unfortunately}, psychological traits like one's personality are high-level and hidden in the innermost corner of data, which is intractable to be uncovered by traditional data mining approaches; The data quality of online social networks is far from sufficient to support such profound psychological analysis, because user behavior records and their attributes are usually very fragmented, missing lots of key information to understand a person in depth; In addition, the social environments in online networks are very complicated, making the interaction patterns between users and their environments underexplored.

In light of these problems, this paper proposes a sociological analysis framework for one's personality in an environment-based view instead of individual-level data mining. Specifically, to comprehensively understand an individual's behavior from low-quality records, we leverage the powerful associative ability of LLMs by designing an effective prompt. In this way, LLMs can integrate various scattered information with their external knowledge to generate higher-quality profiles, which can significantly improve the personality analysis performance. To explore the interactive mechanism behind the users and their online environments, we design an effective hypergraph neural network where the hypergraph nodes are users and the hyperedges in the hypergraph are social environments. We offer a useful dataset with user profile data, personality traits, and several detected environments from the real-world social platform. To the best of our knowledge, this is the first network-based dataset containing both hypergraph structure and social information, which could push forward future research in this area further. By employing the framework on this dataset, we can effectively capture the nuances of individual personalities and their online behaviors, leading to a deeper understanding of human interactions in the digital world. Our code and dataset are open accessible\footnote{\url{https://anonymous.4open.science/r/LLM-HGNN-MBTI-DCC7}
}.

\end{abstract}


\maketitle
\section{Introduction}\label{sec:intro}

Personality online is a crucial aspect that shapes individuals' perceptions, decisions, and social interactions, which has been deeply involved in many online applications such as recommendation systems \cite{wei2017beyond}, online mental treatment \cite{ngantcha2021patient}, and sentiment analysis\cite{zhao2018personality}.

Despite its importance, detecting and accurately analyzing personality traits remains a significant challenge due to their complex and high-level nature. Recently, some works are trying to analyze one's personality from user-generated texts \cite{sun2018personality, sun2020group, mehta2019recent}, images \cite{biel2012facetube, gurpinar2016combining}, audio \cite{principi2019effect}, or interview videos \cite{sun2022your}. Although some progress has been achieved, these individual-level methods still fall short in this regard, struggling to uncover the deep-seated and nuanced aspects of personality from available data, ignoring the implicit networks formed by personality traits, behaviors, and social environments.

\begin{figure}[t]
  \centering 
  \subfloat[Individual-level Analysis]{
    \label{fig:l1}
    \includegraphics[width=0.22\textwidth]{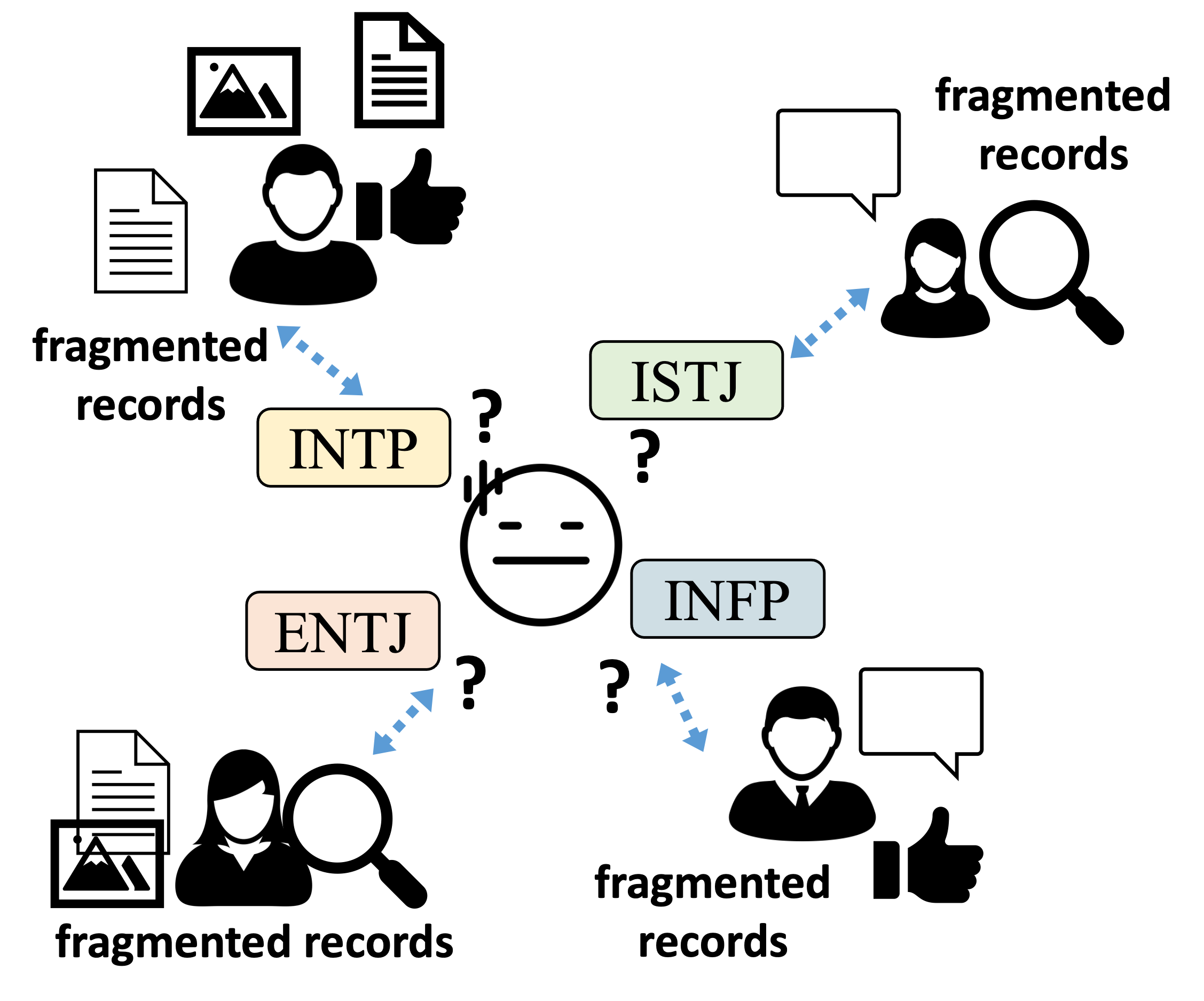}}
  \subfloat[Sociological Analysis]{
    \label{fig:l2}
    \includegraphics[width=0.25\textwidth]{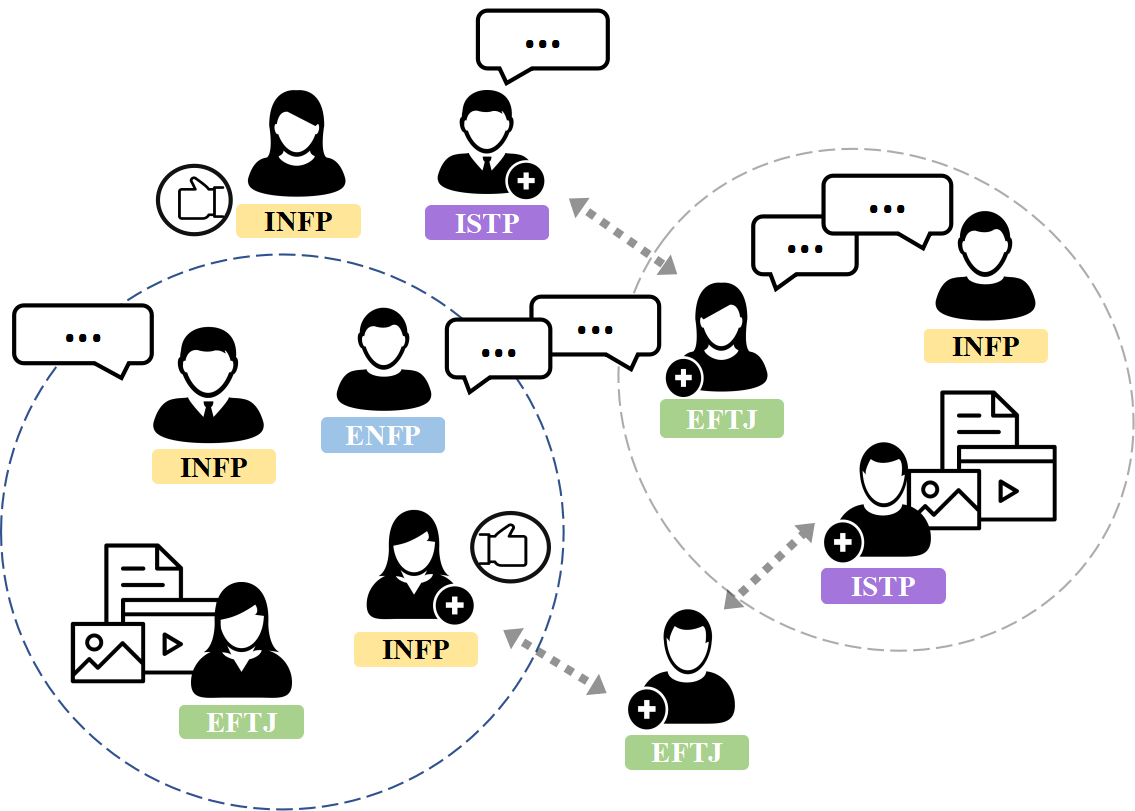}
  }
  \caption{Traditional Approach v.s. Sociological Approach}
  \label{fig:intro1}
\end{figure}

Compared with these individual-level approaches, many sociological and psychological theories \cite{larsen2005personality} have found that users' personality traits, social environments, and behaviors are three closely correlated factors, which are promising to support a more thorough understanding of humans in the digital world. 
To this end, this paper proposes a comprehensive view based on sociological studies in a complex online social network. Compared with traditional individual-level approaches (as shown in Figure \ref{fig:intro1}), our sociological framework provides a more comprehensive understanding of a person, integrating their personality traits, behavior records, and social environment influences.


However, achieving the above vision is never easy. There are at least threefold challenges: 
\begin{itemize}
    \item \textbf{\textit{Challenge 1:}} The first challenge is the scarcity of datasets that support deeper analysis. Most available datasets consist only of text, images, or videos at the individual level. To conduct a more profound analysis, we need comprehensive data that includes social environments, user profiles, behavior records, and crucially, personality labels. These are rarely present in existing datasets due to the high cost of annotation.
    
    \item \textbf{\textit{Challenge 2:}} The second challenge lies in identifying various social environments online and effectively modeling sociological interactions. Online social networks are intricate, with diverse and complicated environments influencing user behavior. Accurately mapping these environments and understanding their impact on individuals require advanced modeling techniques that can capture the complexity of social interactions.

    \item \textbf{\textit{Challenge 3:}} The third challenge is the deep understanding of a person from their online records, where observed attributes are often incomplete and low-quality. User data in social networks are typically fragmented, with missing or irrelevant information making it difficult to construct a coherent and comprehensive profile of an individual’s personality.
\end{itemize}


\textbf{\textit{Presented Work:}} 
In this paper, we propose a novel framework that combines hypergraph neural networks with large language models (LLMs) to address the intricate challenges of analyzing personality traits, behaviors, and social environments in online settings. Our framework stands out for its ability to integrate diverse and fragmented user data into a coherent profile, leveraging the associative power of LLMs and the structural insights provided by hypergraphs. This dual approach allows us to capture the nuanced interactions between individuals and their social environments, offering a deeper understanding of personalities.

Specifically, to address \textbf{\textit{Challenge 1}}, we offer a unique dataset from a burgeoning social platform, which includes user profiles, personality traits, and detailed social environment information. This dataset is the first of its kind to integrate a hypergraph structure with sociological data, providing a rich resource for in-depth analysis. By compiling this comprehensive dataset, we facilitate a level of analysis that was previously unattainable due to the high cost of data annotation and the fragmented nature of existing data sources. To address \textbf{\textit{Challenge 2}}, we develop a hypergraph neural network framework to model the sociological interactions within online environments. This framework utilizes hyperedges to represent complex social environments and captures the interactions between users and their surroundings. By mapping these relationships in a hypergraph, we can model the multi-faceted influence of online social interactions, going beyond traditional pair-wise relationships to explore richer sociological patterns. To address the \textbf{\textit{Challenge 3}}, we leverage the powerful associative capabilities of large language models (LLMs). We design effective prompts that guide the LLMs to integrate scattered and fragmented user information, enhancing the quality and coherence of the generated user profiles. This approach allows us to fill in the gaps in the data, creating a more complete and accurate representation of an individual's personality and behavior.

\textit{\textbf{Contributions:}}
\vspace{-0.5em}
\begin{itemize}
\item Comprehensive Dataset: We provide a unique dataset that includes user profiles, personality traits, and social environments, enabling deeper analysis of online social interactions.

\item Effective Prompting Framework: We design a novel prompting framework that utilizes large language models to integrate and enhance fragmented user data, significantly improving the quality of personality analysis.

\item Innovative Modeling Framework: We develop a hypergraph neural network framework to model the complex sociological interactions within online environments, capturing the nuances of social dynamics and personality influences.

\item Extensive Experimental Validation: We conduct extensive experiments to validate the effectiveness of our framework, demonstrating its ability to accurately analyze and understand individual personalities in online social networks.

\end{itemize}

\section{Preliminaries}\label{sec:pre}
In this section, we briefly introduce the necessary concepts and formulate our target problem.

\noindent \ding{182} \textbf{Personality Traits:}
Personality traits refer to enduring patterns of thoughts, feelings, and behaviors that distinguish individuals from one another. These traits are consistent over time and across situations, contributing to the uniqueness of each person. In the context of psychology, various models have been developed to classify and understand these traits. Two prominent models are the Myers-Briggs Type Indicator (MBTI) and the Enneagram of Personality.

\begin{figure*}[ht]
\centering
\includegraphics[width=0.99\textwidth]{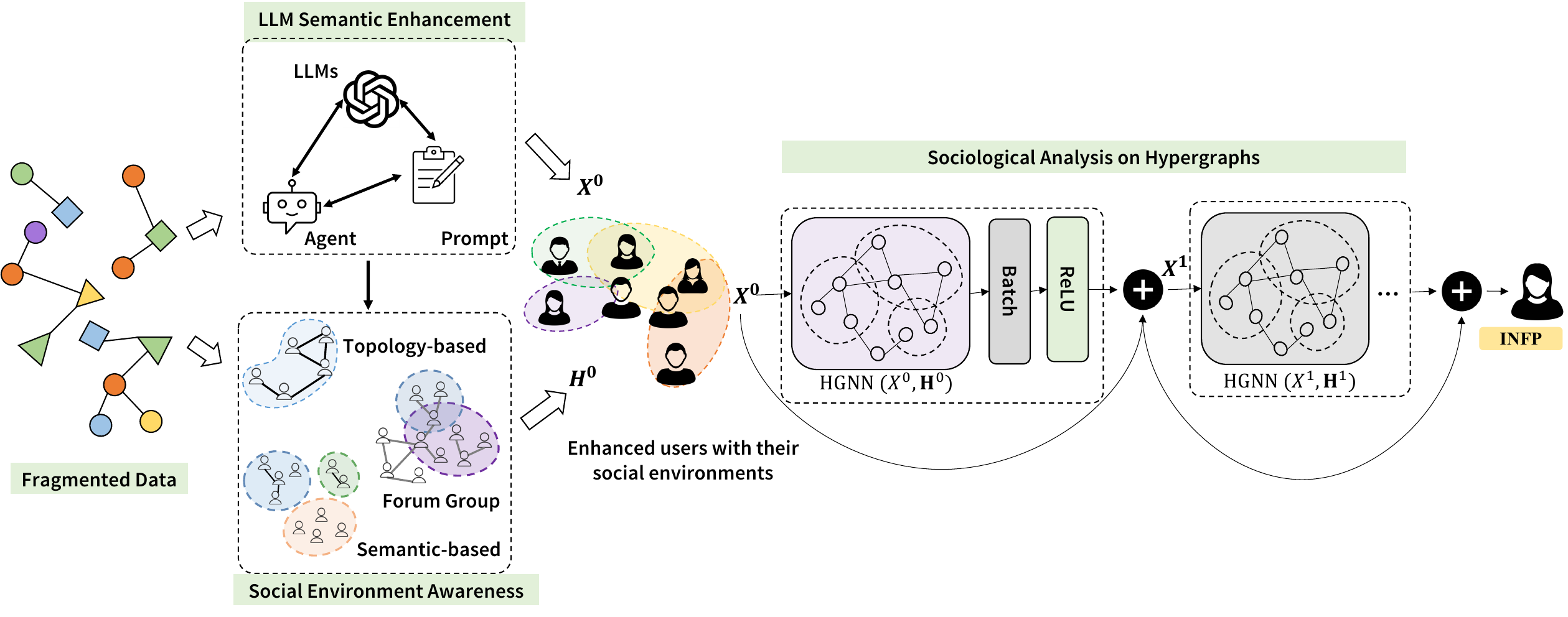}
\caption{The Framework of Our Sociological Personality Analysis. An agent reads the fragmented data and injects the raw information into a carefully designed prompt, which will be used to interact with an LLM. The enhanced content and three kinds of social environments are sent into a hypergraph neural network for sociological personality analysis.}
\label{fig:model}
\end{figure*}

\noindent \ding{183} \textbf{The MBTI} is based on Carl Jung's theory of psychological types \cite{jung2016psychological} and categorizes personality into 16 distinct types across four dichotomies: 
\begin{itemize}
    \item \uline{\textbf{E}}xtraversion / \uline{\textbf{I}}ntroversion, which reflects the source and direction of an individual's energy expression.
    \item \uline{\textbf{S}}ensing / \uline{\textbf{I}}ntuition, which relates to how individuals perceive information.
    \item \uline{\textbf{T}}hinking / \uline{\textbf{F}}eeling, which describes how individuals process information and make decisions.
    \item \uline{\textbf{J}}udging / \uline{\textbf{P}}erceiving, which pertains to how individuals interact with the external world and manage their lifestyle.
\end{itemize}
Each individual’s personality type is denoted by a four-letter code, such as ENFP or ISTJ. Mathematically, we can represent an individual's MBTI type as a vector in a 4-dimensional binary space:

\vspace{-1.3em}
\begin{equation}\label{equ:mbti1}
    \mathbf{T} = (t_1, t_2, t_3, t_4)
\end{equation}
\vspace{-1.3em}

\noindent where each \( t_i \) is a binary variable representing one of the dichotomies, for example: $t_1 \in \{E, I\}, t_2 \in \{S, N\}, t_3 \in \{T, F\}, t_4 \in \{J, P\} $. For computational purposes, we can map these dichotomies to binary values, such as: $t_1 = E \Rightarrow 0, t_1 = I \Rightarrow 1; t_2 = S \Rightarrow 0, t_2 = N \Rightarrow 1; t_3 = T \Rightarrow 0, t_3 = F \Rightarrow 1; t_4 = J \Rightarrow 0, t_4 = P \Rightarrow 1$. Thus, an MBTI type like INFP would be represented as $\mathbf{T} = (1, 1, 1, 1)$. 

On the other hand, since these four dichotomies generate 16 distinct types, we can also define the MBTI type as a unit vector in a 16-dimensional space like: 

\vspace{-1.3em}
\begin{equation}\label{equ:mbti2}
\mathbf{T}_{MBTI}=(p_1, p_2, \cdots, p_{16})
\end{equation}
\vspace{-1.3em}

\noindent where each $p_i$ is a binary variable indicating the presence ($1$) or absence ($0$) of a particular MBTI type. Typically, one type is dominant, resulting in a sparse vector with a single element set to $1$ and the rest to $0$. For example, an individual of ``type \#1'' would be represented as:
$\mathbf{T}_{MBTI} = (1, 0, 0, \cdots, 0)$

\noindent \ding{184} \textbf{Enneagram of Personality} describes 9 distinct personality types and their interrelationships. Following a similar way as MBTI, we can mathematically represent an individual's Enneagram type as a unit vector in a 9-dimensional space:

\vspace{-1.3em}
\begin{equation}\label{equ:enneagram}
    \mathbf{T}_{Enneagram} = (p_1, p_2, \ldots, p_9)
\end{equation}
\vspace{-1.3em}

\noindent\textbf{Note:} The prediction of MBTI traits can be formulaized as \textit{multi-label classification task} with the label format defined by Equation (\ref{equ:mbti1}), or \textit{multi-class classification task} with the label format defined by Equation (\ref{equ:mbti2}). But the prediction of Enneagram types can be formulated as \textit{multi-class classification task} only. For the consistent concern, this paper formulates these two personality predictions as \textit{multi-class classification tasks}.



\noindent \ding{185} \textbf{Hypergraph: }
We define a hypergraph as: $\mathcal{G}=\{\mathcal{V},\mathcal{E},\boldsymbol{X},\boldsymbol{W},\boldsymbol{U}\}$, where $\mathcal{V}$ represents the set of nodes within $\mathcal{G}$. $\mathcal{E}$ is the collection of hyperedges in the hypergraph. Unlike traditional dyadic graphs where each edge connects only two nodes, the edge in a hypergraph (a.k.a hyperedge) can connect more nodes, which means each hyperedge can be represented as a subset of the total nodes $\mathcal{V}$. $\boldsymbol{X} \in \mathbb{R}^{|\mathcal{V}|\times d}$ is the feature matrix of nodes. We use the diagonal entries of two diagonal matrices, $\boldsymbol{W} \in \mathbb{R}^{|\mathcal{E}|\times |\mathcal{E}|}$ and $\boldsymbol{U}\in \mathbb{R}^{|\mathcal{V}|\times |\mathcal{V}|}$, to denote the importance of each hyperedge and each node, respectively.
In this paper, we study each online user as a node in a hypergraph and treat each social environment as a group of users with close social patterns (e.g. connectivity, similar attributes, etc), each of which can be denoted as a hyperedge. The hyperedge weights ($\boldsymbol{W}$) and user weights ($\boldsymbol{U}$) are preset. 

\ding{168} \textbf{Objective: }
Our target problem can be formulated as a node multi-class classification task on a user hypergraph. Specifically, given a user hypergraph $\mathcal{G}$ as defined above, we split user set $\mathcal{V}$ as two disjoint subsets, $\mathcal{V}^{*}$ and $\mathcal{V}^{\circ}$, for learning and predicting, respectively, where $\mathcal{V}^{*}\cup\mathcal{V}^{\circ}=\mathcal{V}$ and $\mathcal{V}^{*}\cap\mathcal{V}^{\circ}=\varnothing$.
The personality labels of $\mathcal{V}^{*}$ can be observed and denoted by $\mathbf{Y}^{*}\in \{0,1\}^{|\mathcal{V}^{*}|\times P}$ where $P$ is the total number of personality types. The predicted personality labels of $\mathcal{V}^{\circ}$ is denoted by $\mathbf{Y}^{\circ}\in \{0,1\}^{|\mathcal{V}^{\circ}|\times P}$ and invisible during model learning. We wish to learn a hypergraph neural network using $\mathcal{G}$ and $\mathbf{Y}^{*}$, and use the learned hypergraph model to achieve better prediction results for $\mathbf{Y}^{\circ}$.


\section{Sociological Personality Analysis}


\textbf{Overview:} Our proposed sociological personality analysis framework is shown in Figure \ref{fig:model}. Unlike existing works that mostly predict one's personality from his/her individual-level and fragmented data, we here propose a new approach, sociological personality analysis, which means we wish to learn in depth the underlying patterns among one's intrinsic sociological traits (e.g. personality), his/her behaviors, and social environmental influence. However, social environments in the real world involve more than just pair-wise connections. This complexity presents significant challenges for conventional dyadic graph models, which often struggle with higher-level complex social interaction modeling. To this end, we go beyond pair-wise relations with hypergraphs and propose a novel deep hypergraph framework utilizing three types of hyperedges (topological neighbors, semantic neighbors, and interest groups) to simulate real-world social environments. In addition, the implications of people's personalities are often hidden in the intricate processing of such complex datasets. To address this issue,  we further design an effective prompt framework to bridge the powerful knowledge from LLMs and users' fragmented records to generate cohesive paragraphs in enhancing the semantic richness and information density. The sociological personality analysis framework has a solid theoretical basis from massive psychological and sociological studies, making our approach more natural to human beings and may serve better for human-centered applications. Next, we present the main component of this framework in detail.

\subsection{Social Environment Awareness}\label{subsec:env}

As mentioned in section \ref{sec:pre}, we define the social environment as a group of people who share some similar characteristics and close social distance in online social networks. We define three kinds of hyperedges to represent mostly used environments as follows:

\begin{itemize}
    \item \textbf{Topology-based Hyperedges:} This set includes hyperedges formed by the k-hop neighbors of a node based on the social network connectivity. Let $\mathcal{N}^k_v=\{u \in \mathcal{V} \setminus \{v\} \mid \operatorname{dist}(v, u) \leq k\}$ be node $v$'s k-hop neighbors where $\operatorname{dist}(v, u)$ denotes the distance between node $v$ and node $u$, then we can formulate topology-based hyperedges as $\mathcal{E}_{top} = \{\{v\}\cup \mathcal{N}^k_v \mid \forall v \in \mathcal{V}\}$. We configure $k = 2$ to encapsulate mutual or one-on-one relationships, reflecting immediate social circles that might influence an individual’s behavior or preferences. By emphasizing local connectivity patterns, this type of hyperedge enhances our understanding of the close links around individuals.
    \item \textbf{Semantic-based Hyperedges}: Beyond topological neighbors, we also explore user features and find the $k$-nearest neighbors of node $v$ based on the feature similarity. For example, we can define the hyperedge set by $\mathcal{E}_{sem}=\{\{v\}\cup \operatorname{KNN}(v) \mid  \forall v \in \mathcal{V}\}$ where $\operatorname{KNN}(v)$ denotes the function that return $K$-nearest neighbors of $v$ based on the feature similarity. This method groups nodes into hyperedges based on shared attributes or behaviors, enriching the hypergraph with detailed relational information.
    \item \textbf{Forum Group-Based Hyperedges}: In many online social platforms, there are many natural interest groups. For example, a topic group (usually denoted by a hashtag) on Twitter, and a discussion group on Weibo, etc. These forum groups can well represent some important social influence on an individual. We here treat such a group as a hyperedge, then we can easily translate the complex group memberships into a structured hypergraph format.
\end{itemize}





\subsection{Semantic Enhancement by LLMs}\label{subsec:llms}


In the real world, the observed user online records are usually low-quality for such thorough analysis of one's psychological traits because there may exist many missing items, incompatible features, and fragmented information. For example, user features often manifest in various forms, such as behavioral items (e.g. like or dislike), continuous values (e.g. account usage time), discrete values (e.g. gender), linguistic cues (e.g. ``about me'' column in one's homepage), and tags content to present their preferences, leading to fragmented datasets.

The low quality of user data in social networks presents significant challenges in creating coherent user understanding and training effective psychological analysis models. Traditional solutions usually heavily rely on human power to clean these data through complicated engineering skills but achieve less improvement. For example, for some missing attributes, they usually set them as preset values without any knowledge; for different attribute formats, they have to develop multiple channels to deal with each kind of attribute separately. All these endeavors seem to be too bloated, leading to unsatisfactory performance.

\begin{figure}[ht]
\centering
\includegraphics[width=0.45\textwidth]{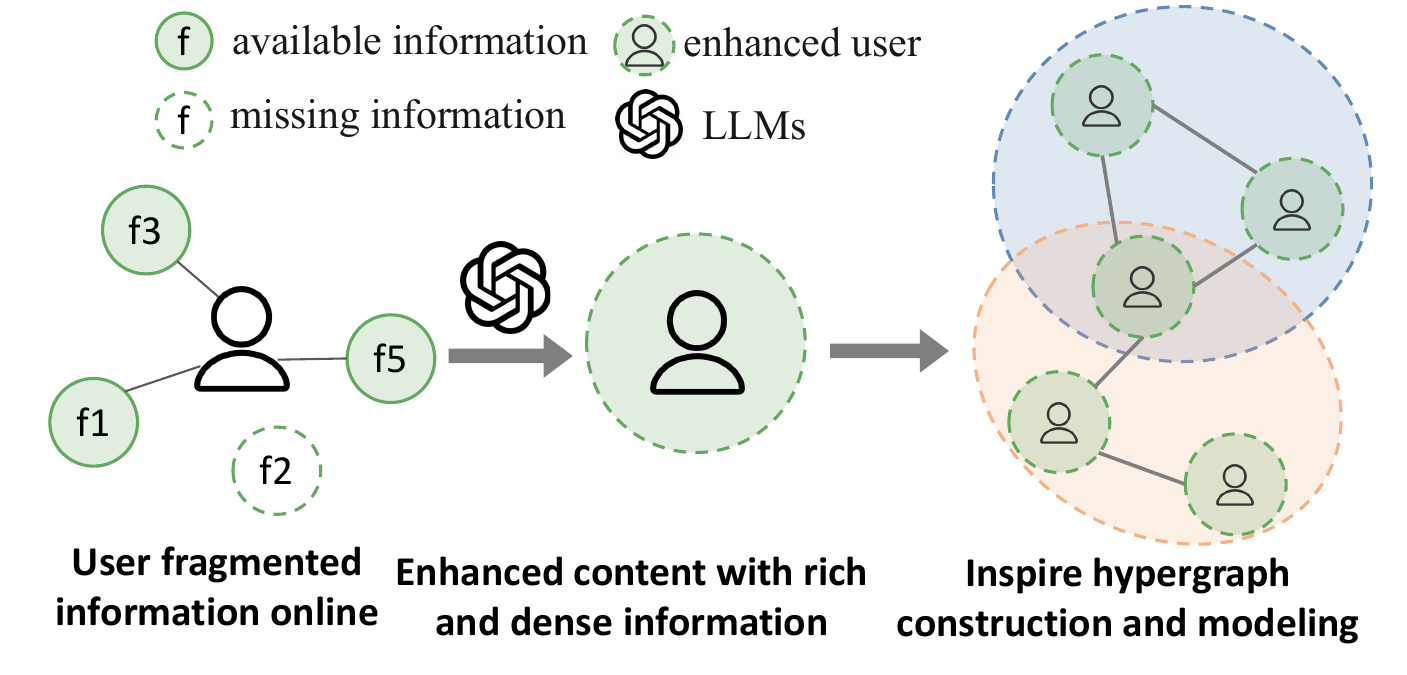}\vspace{-1.2em}
\caption{The Process of Enhancing User Semantic Description with Fragmented Information.}\vspace{-1.2em}
\label{fig:llm_process}
\end{figure}

\textbf{Our Solution.} We have realized that an ideal solution to the above issue is to compress all these attributes in one way, leverage external knowledge to associate the missing item and generate a more comprehensive understanding of users with higher information density. To this end, we utilized three pre-trained Large Language Models (LLMs): GPT-3.5-turbo\footnote{\url{https://platform.openai.com/docs/models}}, LLAMA-2-7b\footnote{\url{https://huggingface.co/meta-llama/Llama-2-7b-chat-hf}}, and gemma\footnote{\url{https://huggingface.co/google/gemma-7b}} to generate descriptive narratives from fragmented user data. The workflow is indicated in Figure \ref{fig:llm_process}. We crafted novel prompts that incorporate attributes or features across user behaviors. LLMs facilitate missing information with preserved knowledge and integrate all characteristics and data of each user, which significantly enhances the efficiency of data processing and model learning. 

With the above idea in mind, the next questions are how to design effective prompts to connect fragmented data to the LLMs' capability, and how to generate a computable feature matrix with high information density:

\ding{182} \textbf{Prompt Design:} In our collected dataset (see section \ref{sec:data} for more detailed information), we first hide the observed personality traits to avoid the risk of label leaking. Then we organize user records as a JSON file that contains many ``(attribute name, attribute content)'' pairs. In this paper, the considered attributes include some key personal and social metrics such as gender, geographical location, self-description, occupation, relationship status, number of followers, community affiliations, etc., which were collected in our raw dataset. With this data, we offer our designed prompt as follows:
\begin{tcolorbox}[title=Our Prompt]
\textbf{Task:} Generate a descriptive paragraph about the given user's persona. 

\textbf{Demand:} \#1. For some unknown attributes, you should try to complete them with your knowledge. \#2. Your description of this user should focus on personal traits, with references to demographic details given below. \#3. Make the description concise and brief. 

\textbf{User Records:} $<$\textit{Attribute Name}$>: <$\textit{Attribute Content}$>;$ $<$\textit{Attribute Name}$>: <$\textit{Attribute Content}$>;\cdots$
\end{tcolorbox}

Intuitively, demand \#1 encourages the LLM to fill in the missing information based on its preserved knowledge and the powerful associative ability; demand \#2 pushes the LLM to ignore irrelevant information and filter out the most useful information for the personality task; and demand \#3 controls divergent thinking of the LLM to avoid the deviation from the original data. We use the above prompt format for a specific user in our dataset and detail the ``User Records'' as: 
``<Name>: <Veritalor Splendy\footnote{For privacy concerns, we present a fake name here. \label{footnote5}}>;  <Gender>: <Female>; <Sexual>: <bisexual>; <Age>: <\textbf{Unknown}>; <About>: <I like to walk and read on weekends, I am a very open person, my ex-girlfriend ...>; <Groups>: <book, walk, ...>...'', then we send the synthetic prompt to a LLM and get the following response:

\begin{tcolorbox}[ title=Example]
Veritalor Splendy\textsuperscript{\ref{footnote5}} is a vibrant and dynamic individual who embodies openness and curiosity. As a female identifying as bisexual, she embraces diversity and inclusivity in her life. While her exact age is unknown, her enthusiasm for weekend walks and reading suggests a \textbf{youthful and energetic spirit}. Veritalor is highly social, actively participating in book and walking groups, which reflects her love for learning and physical activity. Her openness and past relationships indicate a person who values deep connections and is comfortable with her identity. She enjoys exploring new ideas and experiences, making her a well-rounded and engaging personality.
\end{tcolorbox}

By deploying the LLMs, we enhance the dataset with both demographic details and descriptive aspects of user social interactions and personal identity based on their fragmented records. The process is capable of weaving fragmented pieces of information into cohesive narratives, capturing the essence of user personalities, interests, and behaviors in acontextually rich paragraph.

\ding{183} \textbf{Semantic Encoding:} 
After the enhanced content is generated for each user, we input the list of generated descriptions into the Sentence Transformer \cite{Reimers2019} to derive vectorized embeddings like $\mathbf{X}_{llm}\in \mathbb{R}^{N\times d}$ where $N$ is the total number of users and $d$ is the embedding dimension number. These embeddings are then integrated into the node features for generating the hypergraph network, and each input node is represented by contextually rich feature vectors, allowing our model to identify and learn the nuanced patterns associated with various personality types more effectively.

\subsection{Personality Analysis via Hypergraph Model}


Inspired by many sociological studies, our sociological personality analysis framework stands out from traditional individual-level analysis by carefully modeling the complicated interaction between users and their social environments. To this end, we leverage a hypergraph neural network with skip connection skills to model such complex interactions.

Let $\boldsymbol{H}\in \{0,1\}^{|\mathcal{V}| \times |\mathcal{E}|}$ denote whether a node is present in a given hyperedge. For example,  $\boldsymbol{H}_{uk}=1$ means the $u$-th user of $\mathcal{V}$ is the member of the $k$-th environment (hyperedge), and $\boldsymbol{H}_{uk}=0$ for the opposite. The  degree of $k$-th hyperedge is defined as $d^e_k=\sum_{u \in \mathcal{V}}\boldsymbol{U}_u \boldsymbol{H}_{uk}$, leading to a diagonal hyperedge degree matrix as  $\boldsymbol{D}^e=\boldsymbol{Diag}(d^e_1, \cdots, d^e_{|\mathcal{E}|}) \in \mathbb{R}^{|\mathcal{E}| \times |\mathcal{E}|}$. The node degree matrix can be also given similarly like $\boldsymbol{D}^v=\boldsymbol{Diag}(d^v_1, \cdots, d^v_{|\mathcal{V}|}) \in \mathbb{R}^{|\mathcal{V}| \times |\mathcal{V}|}$ where $d^v_k=\sum_{k \in \mathcal{E}^h} \boldsymbol{W}_k \boldsymbol{H}_{uk}$. Then the hypergraph Laplacian matrix can be defined as follows:
\begin{equation}
\boldsymbol{\Delta}=\boldsymbol{I}-\left(\boldsymbol{D}^v\right)^{-\frac{1}{2}} \boldsymbol{U} \boldsymbol{H} \boldsymbol{W}\left(\boldsymbol{D}^e\right)^{-1} \boldsymbol{H}^{\mathrm{T}} \boldsymbol{U}\left(\boldsymbol{D}^v\right)^{-\frac{1}{2}}
\end{equation}

We are particularly interested in the second item of the above equation, which is $\left(\boldsymbol{D}^v\right)^{-\frac{1}{2}} \boldsymbol{U} \boldsymbol{H} \boldsymbol{W}\left(\boldsymbol{D}^e\right)^{-1} \boldsymbol{H}^{\mathrm{T}} \boldsymbol{U}\left(\boldsymbol{D}^v\right)^{-\frac{1}{2}}$, because this matrix reflects that the information flows from nodes to hyperedges and then from hyperedges to nodes. This is very consistent with our observations in the sociological area, which indicates the interaction between users and their environments. Following this intuition, we can update the user input feature matrix as follows:
\begin{equation}
\mathbf{Z}^{\ell+1} = \left(\boldsymbol{D}^v\right)^{-\frac{1}{2}} \boldsymbol{U} \boldsymbol{H} \boldsymbol{W}\left(\boldsymbol{D}^e\right)^{-1} \boldsymbol{H}^{\mathrm{T}} \boldsymbol{U}\left(\boldsymbol{D}^v\right)^{-\frac{1}{2}}\mathbf{X}^{\ell} \boldsymbol{\Theta}
\end{equation}
where $\boldsymbol{\Theta}$ is the layer-specific trainable parameters, $\ell=0,\cdots,L$ denotes the hypergraph neural network layer. When $\ell=0$, $\mathbf{X}^{0}=\mathbf{X}_{llm}$ is the user feature matrix obtained in section \ref{subsec:llms}. To avoid over-smooth risk, we use the skip connection skill \cite{li2020deepergcn, he2016deep} to connect two adjacent hypergraph neural network layers. Here the skip connection skill is followed by:
\begin{equation}
    \mathbf{X}^{\ell+1} =\sigma\left(ReLU(BatchNorm(\mathbf{Z}^{\ell+1})) + \mathbf{X}^{\ell}  \right)
\end{equation}
where $\sigma(\cdot)$ is the activate function such as ReLU.

Considering that personalities naturally have an imbalanced distribution, we use the focal loss function, which incorporates the focusing parameter $\gamma$ to emphasize the correction of misclassified examples and enhance the recognition of minority classes, to improve model robustness against imbalanced data:
\begin{equation}
L = -\frac{1}{N} \sum_{i=1}^n \sum_{c=1}^C w_c \cdot y_{i,c} \cdot (1-\hat{y}{i,c})^\gamma \log(\hat{y}{i,c})
\end{equation}
where $C$ is the number of personality types, $N$ is the number of labeled users. We utilize $w_c$ as the weight to adjust the loss contribution based on the inverse frequency of each personality type.

\section{Data Analysis}\label{sec:data}
\subsection{Awkward Data Support} Although sociological analysis of personality is important, currently most available datasets can not support such a thorough analysis. \textbf{First}, personality labeling needs huge human power, leading to labeled samples far from sufficient in many existing datasets. Therefore, many datasets \cite{sun2022your, sun2020group, zhao2018personality} only contain no more than 1,000 samples, which is far from enough for training effective models. \textbf{Second}, most existing datasets focus on one single modality such as text alone\cite{sun2020group}, video alone \cite{sun2022your}, or multi-channel features \cite{zhu2023personality} at the individual level, but none of them contain social environments and users interaction behaviors. Therefore these datasets can only support individual detection with naive models but are far from enough to explore sociological analysis of one's behavior,  personality, and social environments. \textbf{Third}, even though these datasets contain some personality labels, most of them only contain one kind of personality measurement (such as BMTI). One single personality measurement is less comprehensive and accurate to reflect one's true intrinsic nature.

\subsection{Data Preparation}
To push forward the academic research in sociological personality analysis, we offer a new dataset collected from a vibrant online forum, PersonalityCafe\footnote{\url{https://www.personalitycafe.com}}, from October to December 2023. Many users of this online social platform will conduct multiple personality tests and their personality types will serve as a tag on their homepage. The mainstream life on this platform is built upon such personality tags.

Our dataset\footnote{\url{https://anonymous.4open.science/r/LLM-HGNN-MBTI-DCC7}} includes more than 85,000 users, which is far larger than existing datasets. More than 17,000 of them have at least two kinds of personality labels (BMTI, Enneagram). The dataset contains more than 59, 354 following interactions, more than 200, 000 other kinds of interactions (such as quote, mention, etc.), and sufficient social environments as hyperdges form. Some of the notable features selected for our analysis are included in four categories:
\begin{itemize}
    \item Self-reported basic information: This included basic user details such as location, sexual orientation, gender, about, followers, forum participated, MBTI and Enneagram personality traits. For the whole information, please see our open dataset.
    \item The interaction behaviors: User interactions were analyzed to map the pair-wise network, highlighting influential patterns and clusters. In our dataset, we totally collect more than 200, 000 pair-wise interaction links including following, quoting, mentioning, etc. Based on these links and other information, we supplement higher-level environments with the approaches in section \ref{subsec:env}.
    \item Forum group participation: As a natural social environment in many online social networks, forum groups indicate users' activities and tastes. We collect 341 groups in total. The interactions in various forum groups were also recorded, offering insights into their interests and engagement patterns. 
    \item Self-reported personality traits: Each user's self-reported MBTI/Enneagram type is crucial in aligning their social behaviors with specific personality traits. We here treat the personality traits as user labels. Besides MBTI and Enneagram types, part of these users also offer their BigFive personality types, which is another popular personality labeling system.
\end{itemize}

\subsection{Data Analysis}

\begin{figure}[ht]
\centering
\includegraphics[width=0.38\textwidth]{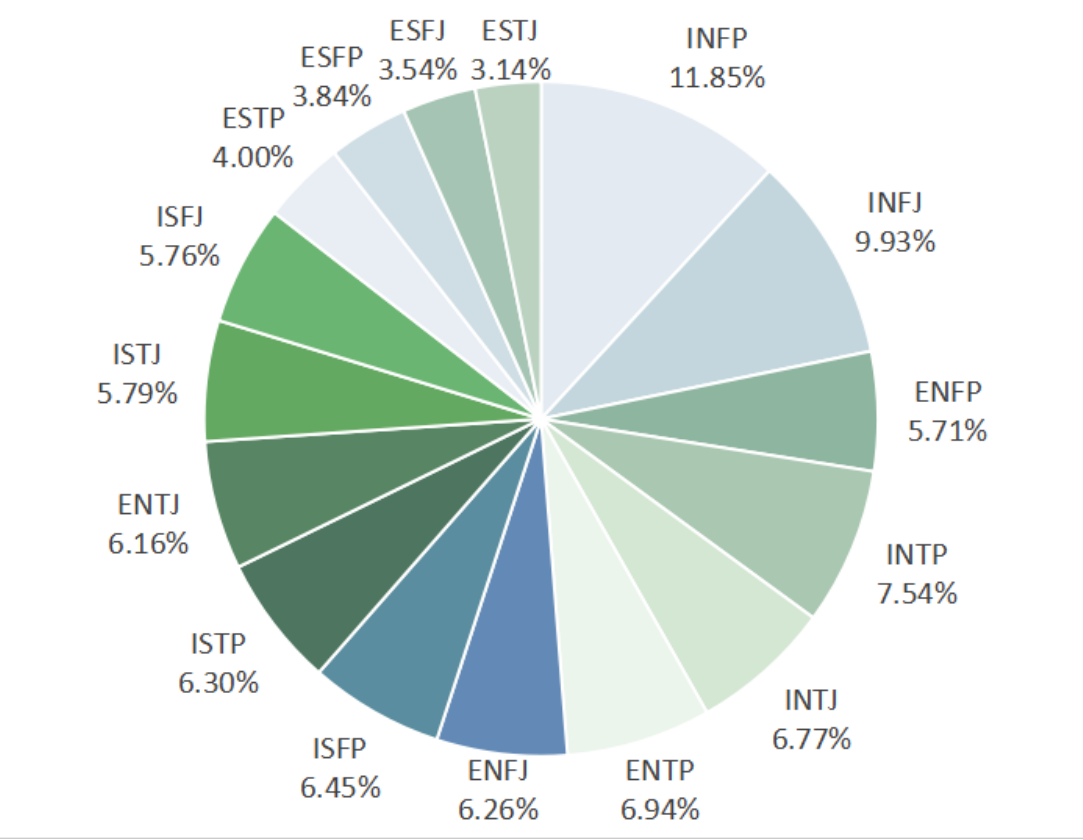}\vspace{-1.2em}
\caption{Personality distribution (MBTI)}\vspace{-1.8em}
\label{fig: MBTI personality}
\end{figure}

\ding{182} \textbf{Personality Distribution:} As shown in Figure \ref{fig: MBTI personality}, certain personality types appear more frequently than others. The most common MBTI type among the users is INFP, with 11.85\% individuals, followed by INFJ with 9.93\%, and INTP with 7.54\%. The least common types are ESTJ and ESFJ, with 3.14\% and 3.54\%, respectively. The unequal distribution of different MBTI personalities online suggests that people with varying personalities engage with and behave differently on the internet. This observation is also confirmed in many related works \cite{alsadhan2017estimating}. \cite{patel2020personality}

\begin{figure}[ht]
\centering
\subfloat[Enneagram v.s. MBTI]{
\label{fig:enn_mbti}
\includegraphics[width=0.25\textwidth]{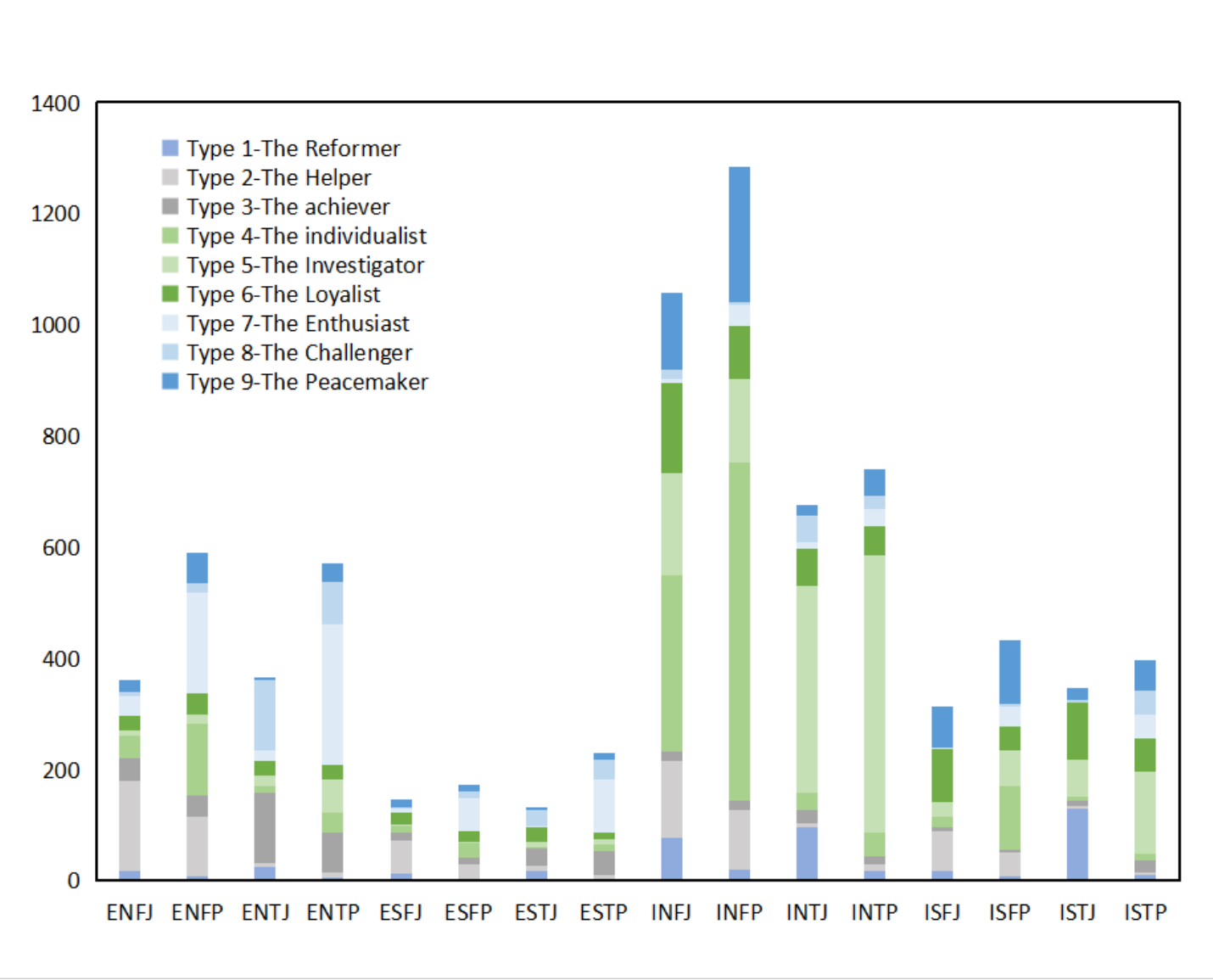}
}
\subfloat[Follower Count v.s. Personality]{
\label{fig:fol_p}
\includegraphics[width=0.235\textwidth]{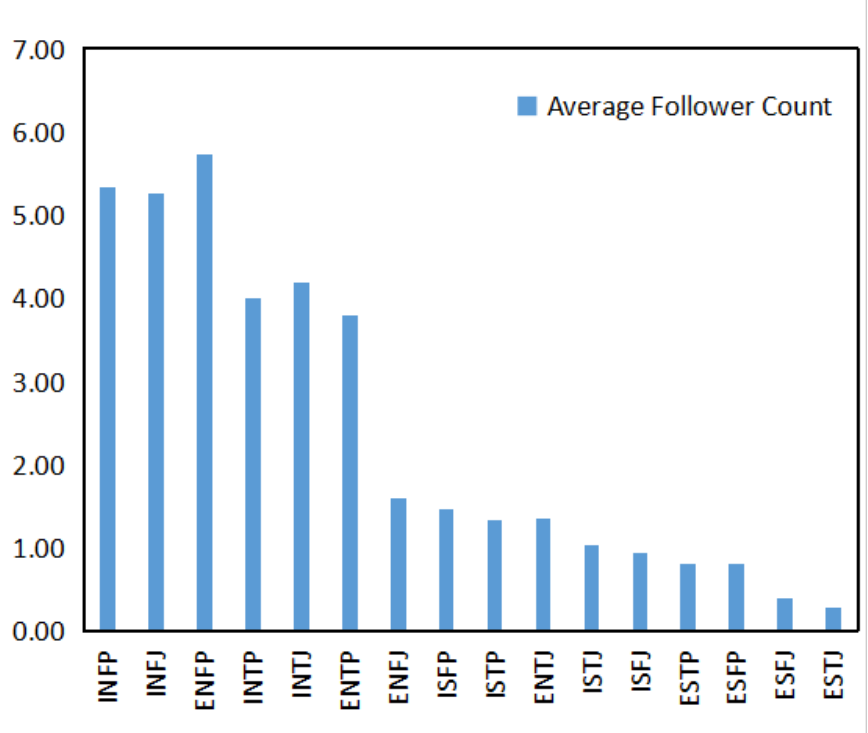}
}\\ \vspace{-1.2em}
\subfloat[Forum Groups v.s. Personality]{
    \label{fig:for_p}
    \includegraphics[width=0.245\textwidth]{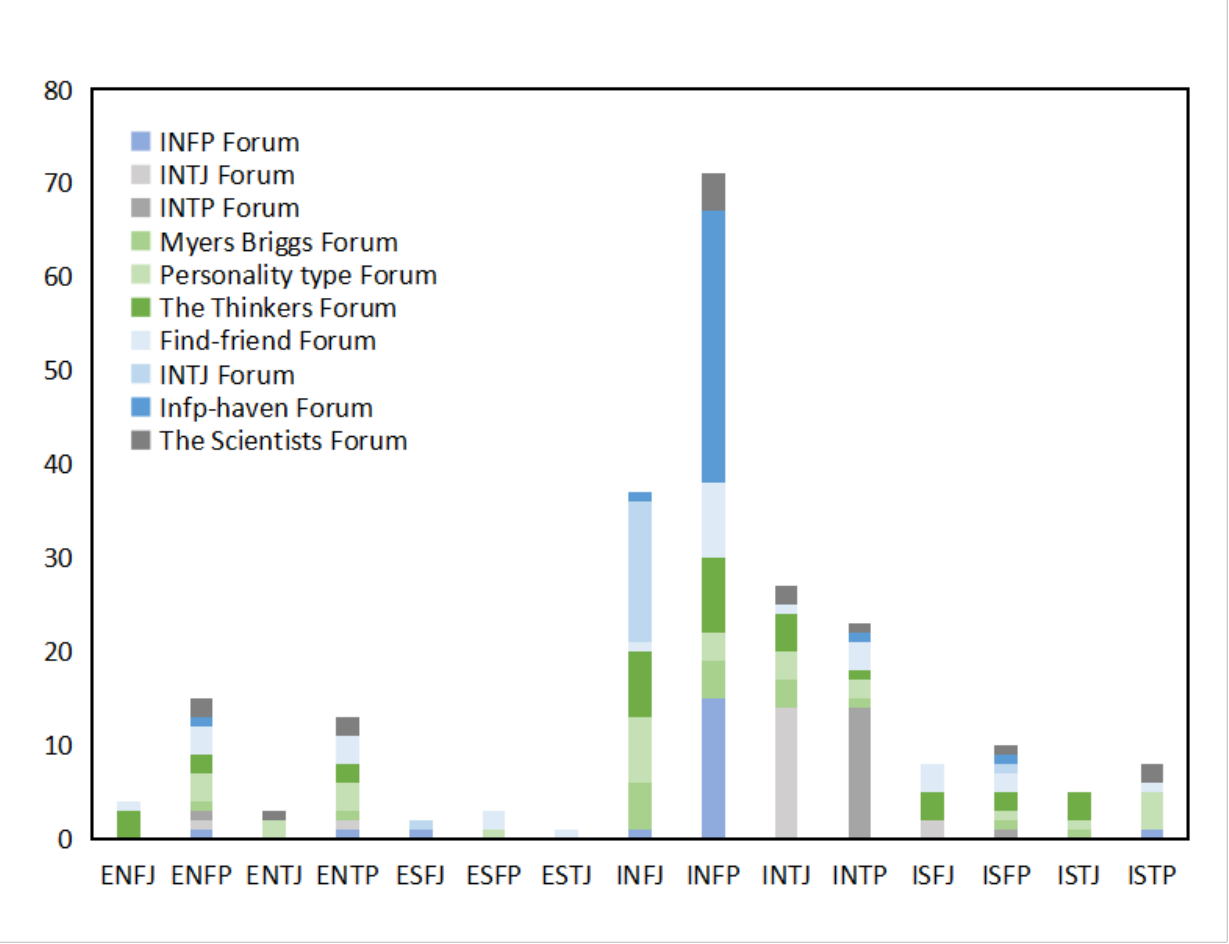}
    }
\subfloat[Gender v.s. Personality]{
\label{fig:gen_p}
\includegraphics[width=0.24\textwidth]{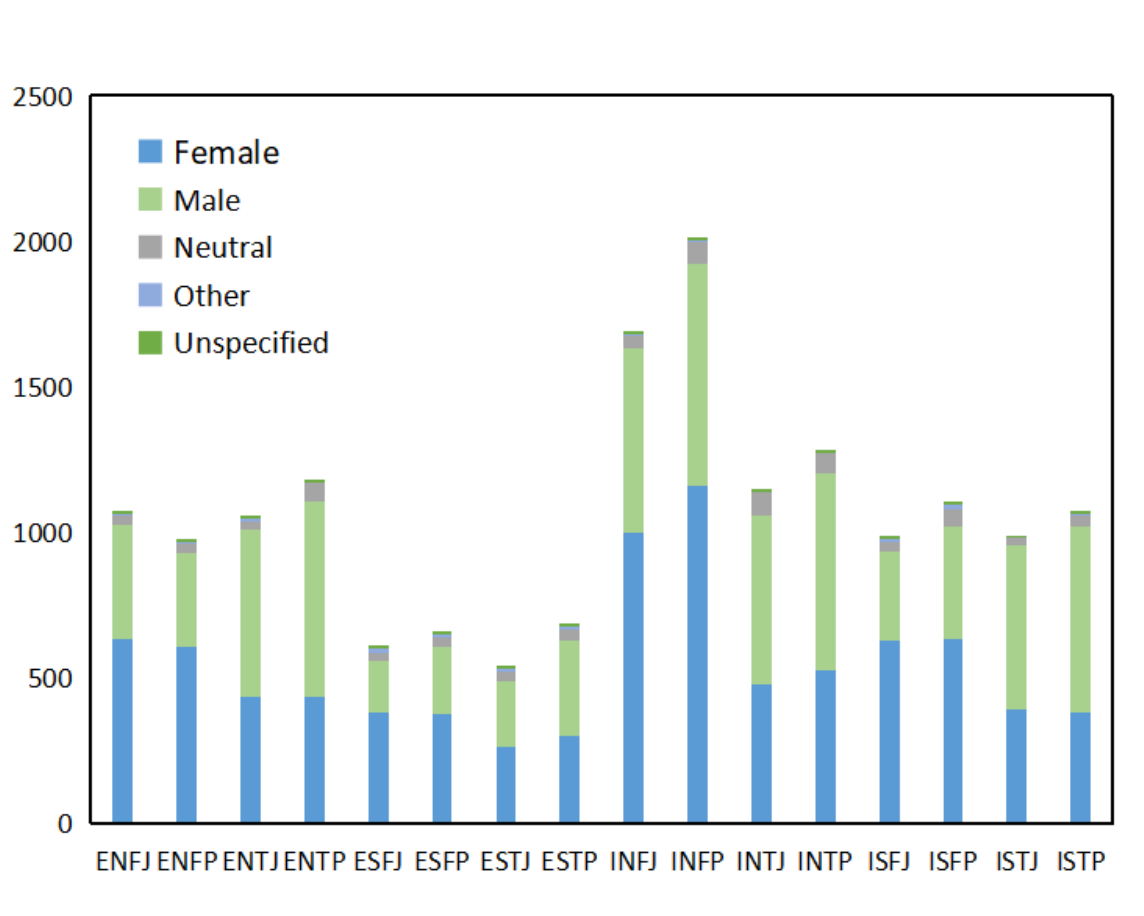}
}\vspace{-1.em}
\caption{Data Distribution Grouped by Personality}
\label{fig:data_sta}
\end{figure}

\noindent \ding{183} \textbf{MBTI and Enneagram:} In Figure \ref{fig:enn_mbti}, we observe the relationships between MBTI types and the Enneagram. For instance, INFP, INFJ, and ENFP display a diversity of Enneagram types. A significant presence of Type 4 (Individualist), Type 5 (Investigator), and Type 9 (Peacemaker) among INFPs suggests a trend toward individualism, intellectualism, and peacemaking. INFJs are also prominently associated with Type 5, implying analytical and contemplative traits. Our findings are also reflected in other related research \cite{Wagner1983Reliability}.

\noindent \ding{184} \textbf{Following v.s. Personality:} An interesting finding from Figure \ref{fig:fol_p} is that many ``\_NF\_'' format personalities (like INFP, INFJ, ENFP, etc) have higher follower counts. In the MBTI theory, ``\_NF\_'' personalities are described as ``idealist'', which is very consistent with our finding.

\begin{table*}[ht]
\centering
\caption{Overall Performance of MBTI and Enneagram Traits. IMP means the average improvement of our method over all metrics compared with the rest baselines.}
\label{tab:performance}
\resizebox{\textwidth}{!}{%
\begin{tabular}{@{}llllll|lllll@{}}
\toprule
                 & \multicolumn{5}{c|}{LLM Enhanced}               & \multicolumn{5}{c}{Raw Feature as Input}    \\ \midrule
\textbf{MBTI Traits}      & Accuracy & AUC & Macro F1 & Micro F1 & IMP     & Accuracy & AUC & Macro F1 & Micro F1 & IMP     \\\midrule \rowcolor[HTML]{EFEFEF} 
Ours             & \textbf{0.8659}   & \textbf{0.9357}    & \textbf{0.8652}   & \textbf{0.8659}   & -        & \textbf{0.8538}   & \textbf{0.9349}    & \textbf{0.8470}    & \textbf{0.8538}   & -        \\ \rowcolor[HTML]{EFEFEF} 
HGNNP            & 0.8352   & 0.9301    & 0.8187   & 0.8352   & 3.4\%  $ \uparrow $  & 0.8282   & 0.9266    & 0.8236   & 0.8282   & 2.5\% $ \uparrow $   \\
GAT              & 0.7347   & 0.9206    & 0.6787   & 0.6644   & 19.3\% $ \uparrow $   & 0.7641   & 0.9042    & 0.7402   & 0.7325   & 11.5\% $ \uparrow $  \\
Gtransformer     & 0.7051   & 0.9070     & 0.6975   & 0.7051   & 18.2\% $ \uparrow $  & 0.7028   & 0.9057    & 0.6741   & 0.6900     & 18.5\% $ \uparrow $  \\
GCN              & 0.7030    & 0.9071    & 0.6416   & 0.7037   & 21.1\% $ \uparrow $  & 0.7002   & 0.9064    & 0.6438   & 0.7002   & 19.6\% $ \uparrow $  \\\rowcolor[HTML]{EFEFEF} 
Chat-gpt-3.0     & 0.0891   & 0.5056    & 0.0290    & 0.0781   & 1212.3\% $ \uparrow $ & 0.0669   & 0.5015    & 0.0145   & 0.0669   & 2045.1\% $ \uparrow $ \\ \midrule
\textbf{Enneagram Traits} & Accuracy & AUC  & Macro F1 & Micro F1 & IMP     & Accuracy & AUC  & Macro F1 & Micro F1 & IMP     \\\midrule
\rowcolor[HTML]{EFEFEF} Ours             & \textbf{0.9477}   & \textbf{0.5880}     & 0.8652   & \textbf{0.9477}   & -        & \textbf{0.9406}   & 0.5349    & \textbf{0.8926}   & \textbf{0.9406}   & -        \\
\rowcolor[HTML]{EFEFEF} HGNNP            & 0.9454   & 0.5854    & \textbf{0.8676}   & 0.9454   & 0.2\% $ \uparrow $   & 0.9285   & \textbf{0.5837}    & 0.8579   & 0.8727   & 1.2\% $ \uparrow $   \\ 
GAT              & 0.8959   & 0.4943    & 0.6787   & 0.8959   & 14.5\% $ \uparrow $  & 0.7641   & 0.5042    & 0.7402   & 0.7641   & 18.2\% $ \uparrow $  \\
Gtransformer     & 0.8851   & 0.5070     & 0.8975   & 0.8851   & 6.6\% $ \uparrow $   & 0.8828   & 0.5057    & 0.8465   & 0.8828   & 6.1\% $ \uparrow $   \\
GCN              & 0.8830    & 0.5071    & 0.8541   & 0.8830    & 8.0\% $ \uparrow $   & 0.8689   & 0.5064    & 0.6438   & 0.8689   & 15.2\% $ \uparrow $  \\
\rowcolor[HTML]{EFEFEF} Chat-gpt-3.0  & 0.0891   & 0.5867    & 0.0740    & 0.0891   & 749.2\% $ \uparrow $ & 0.0871   & 0.4822    & 0.0328   & 0.0871   & 1148.0\% $ \uparrow $ \\
\bottomrule
\end{tabular}%
}
\end{table*}

\noindent \ding{185} \textbf{Group Interests:} As shown in Figure \ref{fig:for_p}, the group involvement of MBTI types points to distinct preferences. INFPs are notably active in Myers-Briggs-related forums, while INTPs participate in groups focusing on intellectual debates like ``The Thinkers'' and ``The Scientists''. INTJs' involvement in ``find-friend'', although less pronounced, reflects a willingness to form connections beyond their intellectual interests.

\noindent \ding{186} \textbf{Gender Difference:} From Figure \ref{fig:gen_p}, we observe some differences in MBTI traits across males and females. For instance, the INFJ and INFP personality types have a higher proportion of females, while the INTJ and INTP types show a more balanced distribution but still have a significant male presence. The ENTJ and ESTP types also display a notable gender disparity, with more males than females. These observations are consistent with MBTI theory. In specific, INFJ and INFP types are often associated with traits like empathy, intuition, and strong interpersonal skills. Traditionally, these traits have been more encouraged and valued in females due to societal norms and gender roles that emphasize caregiving and emotional intelligence. INTJ and INTP types are characterized by analytical thinking, independence, and a preference for logic over emotions. Historically, these traits have been more encouraged in males due to societal expectations that favor rationality and independence in men. However, the balanced distribution indicates that these traits are not exclusive to one gender and that both males and females can exhibit these characteristics. ENTJ types are known for their leadership skills, assertiveness, and strategic thinking, while ESTP types are associated with being energetic, pragmatic, and action-oriented. These traits are often encouraged in males, who are traditionally expected to take on leadership roles and be more assertive.

\begin{figure}[h]
\centering
\subfloat[Follower Count]{
    \label{fig:power_fol}
\includegraphics[width=0.24\textwidth]{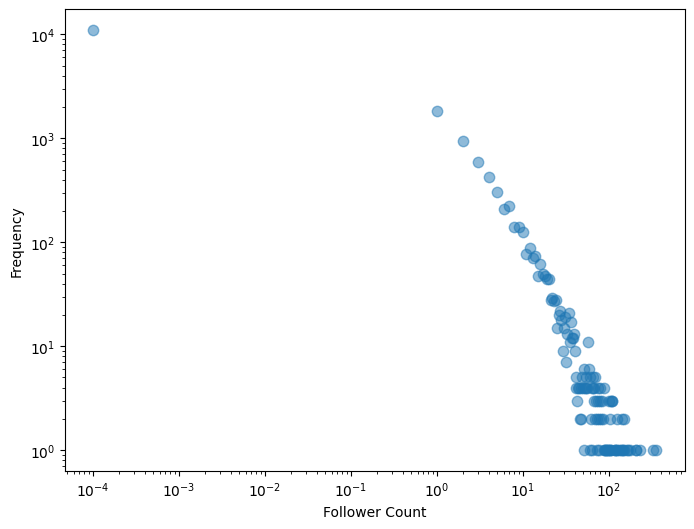}
}
\subfloat[Group Size]{
    \label{fig:l2}
\includegraphics[width=0.24\textwidth]{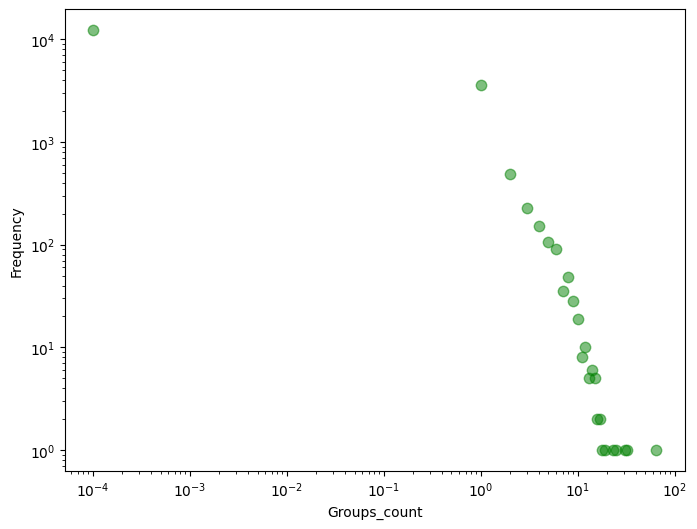}
}\vspace{-1.2em}
\label{fig:power_group}
\caption{Distributions of Follower Count and Group Size}\vspace{-1.5em}
\label{fig:power_law}
\end{figure}

\noindent \ding{187} \textbf{Power-law Analysis} Power-law distribution is one of the most notable patterns in many online social networks and has become a standard to evaluate data rationality. Figure \ref{fig:power_law} from our dataset indicates that most entities have a low follower and group count, with only a few entities having a very high follower count, which is a common phenomenon in social networks.

\section{Experiments}\label{sec:exp}

\subsection{Baselines and Experimental Settings}
We compare our framework with five baselines. Specifically, we have three graph neural networks (GCN \cite{verma2019stability}, GAT \cite{yan2020fpgan}, and Graph Transformer \cite{Yun2019}). We also deploy the large language model (Chat-gpt-3.0) to directly predict the MBTI types based on generated descriptions. Since our framework is compatible with any hypergraph neural network, we also introduce another hypergraph model (HGNNP \cite{gao2022hgnn}) in our framework to see the universality of our approach. We use Adam optimizer with a learning rate of 0.001, weight decay 5e-4. We set max training epochs as 500. Data division was conducted randomly, adhering to an 8:1:1 ratio for training, validation, and testing. The main results are averaged on 5 independent repetitions.


 



\subsection{Overall Performance}
Table \ref{tab:performance} presents the overall performance of various models in detecting MBTI and Enneagram traits. The models are categorized into three groups: (1) our sociological analysis framework, which is also compatible with other hypergraph models like HGNNP; (2) dyadic graph models like GAT, Gtranformer, and GCN; and (3) LLM as the predictor directly. To confirm the contribution of LLMs in solving fragmented data, we use LLMs to generate enhanced features and also study how these models perform with the raw features as input directly.

Overall, our method consistently outperforms other baselines in accuracy, AUC, macro F1, and micro F1 scores. This confirms the effectiveness of our hypergraph-based approach in leveraging sociological analysis within complex social environments to accurately detect personality traits. Specifically, our method and HGNNP, both hypergraph-based models, show superior performance in most metrics. Our method achieves the highest accuracy (0.8659 for MBTI and 0.9477 for Enneagram) and macro F1 scores (0.8652 for MBTI and 0.8652 for Enneagram). This confirms the advantage of hypergraph-based models in capturing complex interactions between users and their social environment. The three graph neural networks (GCN, GAT, and GTransformer) use pair-wise relations. While they perform reasonably well, their scores are generally lower than our method. For instance, GCN achieves an accuracy of 0.7030 for MBTI and 0.8830 for Enneagram, while GAT achieves 0.7347 for MBTI and 0.8959 for Enneagram. These results highlight the limitation of dyadic graphs in fully capturing the nuanced social interactions. The large language model (Chat-gpt-3.0) performs significantly worse across all metrics, with an accuracy of only 0.0891 for both MBTI and Enneagram traits. In the meanwhile, the table also indicates that the compared models with LLM Enhanced features generally perform better than the raw features. These observations indicate that the studied LLM can well deal with fragmented data and improve the data quality, but they are less effective in predicting personality traits directly. With our designed prompt in section \ref{subsec:llms}, we leverage the LLM's strength and avoid its inference shortage in personality analysis.



\subsection{Impact Analysis on LLMs} 
\begin{figure}[ht]
\centering
\vspace{-1.1em}
\includegraphics[width=0.38\textwidth]{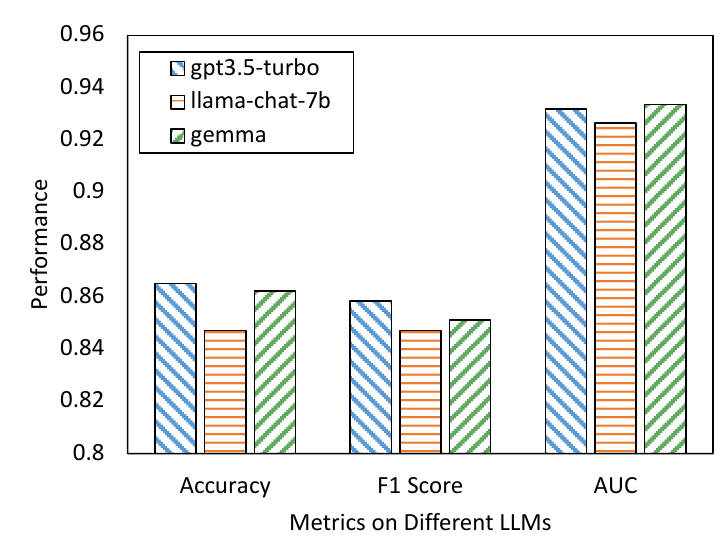}\vspace{-1.4em}
\caption{Effectiveness on LLM Selection}\vspace{-1.em}
\label{fig:abl_llms}
\end{figure}

We systematically deploy three LLMs (gpt3.5-turbo, llama-chat-7b, and gemma) in the same settings to compare user descriptions based on the same user profiles. For all users, we use these three LLMs to generate their content and then we use the content in our framework to see the performance as indicated in Figure \ref{fig:abl_llms}. 

The gpt3.5-turbo presents commendable performance, with an accuracy of 0.8652, which can strongly support our framework. Its AUC was 0.93186, indicating a strong ability to distinguish between classes. Moreover, the F1 Score shows that it maintains a good balance between false positives and false negatives. The gemma model presented an accuracy close to that of gpt3.5-turbo, suggesting a parallel capacity in making correct predictions. Its AUC was the highest among the three at 0.93350, potentially making it the best at distinguishing between classes. 
Llama-chat-7b, with an accuracy of 0.8471, displayed commendable classification abilities, albeit slightly less efficient than gpt3.5-turbo, and still represents a strong performance. However, an intriguing phenomenon was observed exclusively in the llama-chat-7b model; it generated narrative content that was not originally present in the user information. This emergence of data ``illusions—narrative'' elements, generated by the LLM that are not rooted in the input data, was not yet detected in the gpt3.5-turbo or gemma-7b. While we could not quantify the exact number of these generated narrative elements over all the natural language outputs, the specific occurrence within the llama-chat-7b model points to a need for additional investigation. Questions arise as to whether the generation of illusions is influenced by factors such as data sparsity, or the frameworks, training, and fine-tuning procedures specific to each LLM. A case of the generated content is presented in the text boxes below.
\begin{tcolorbox} [title= Original user profile]
<Username: Maki>
<Follower: 0>
<Sexual: Asexual>
<Gender: Female>
<About: Hello. You can call me Maki. I go around the internet with other names, too, so you can also call me Mew or Kikichirin, which I commonly go around by. Nicknames are also fine.>
\end{tcolorbox}
\begin{tcolorbox} [title= GPT-3.5-Turbo]
Maki is a unique individual who identifies as female and embraces asexual identity. She is characterized by an Enneagram type 5 and is from Brazil. She may also go by the names Mew or Kikichirin and is not fluent in English. Despite any language barriers, she is comfortable with the nicknames and is \textbf{open to conversation and connection}. 
\end{tcolorbox}
\begin{tcolorbox} [title = Gemma-7B]
Maki: An introspective and creative soul from Brazil, Maki embraces her individuality and expresses herself through a variety of online personas. Despite her \textbf{professed shyness}, she exudes a \textbf{quiet confidence} and a \textbf{deep appreciation for language and culture}.
\end{tcolorbox}
\begin{tcolorbox} [title = Llama-7B]
Maki - This user is a female who is Brazilian and has a different name in other online communities. She is asexual and has a \textbf{humorous tone in her bio}, indicating a \textbf{light-hearted personality}. She is part of the group "infp-haven."
\end{tcolorbox}

\subsection{Impact Analysis on Social Environments} 

\begin{figure}[ht]
\centering
\includegraphics[width=0.38\textwidth]{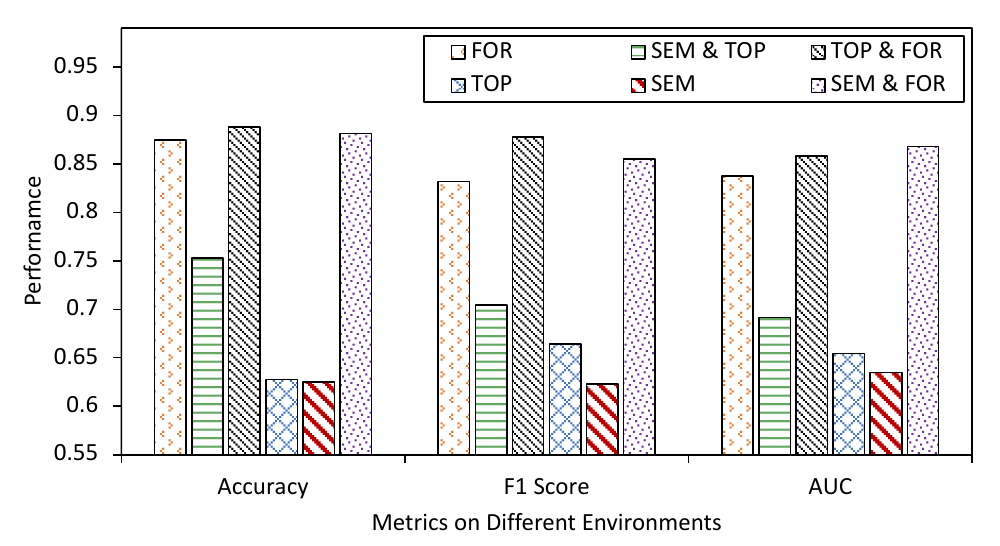}\vspace{-1.em}
\caption{Impact Analysis on Social Environments}\vspace{-1.5em}
\label{fig:abl_hyperedge}
\end{figure}

To delve into the influence of hypergraph construction on model performance, our study explored different hyperedge combinations in the classification process. For simplicity, we use ``FOR'' to denote forum group-based hyperedges as defined in section \ref{subsec:env}, ``TOP'' denotes topology-based hyperedges, and ``SEM'' denotes semantic-based hyperedges. From the results in Figure\ref{fig:abl_hyperedge}, forum groups demonstrate the significance of community structures in predicting personality types. The combination of forum groups and other types of hyperedges also achieved higher performances, which implies that the social groups in online networks are strongly aligned with user personality traits. Moderate performance with ``TOP'' or ``SEM'' alone suggests that they also contribute positively to the model performance. The nuanced performance differences between each configuration suggest that there is no one-size-fits-all approach to hypergraph construction for personality classification. Instead, it’s about finding the right synergy between different types of hyperedges to capture the full spectrum of social behaviors and tendencies reflected in online community data.    

\subsection{Impact Analysis on Label Ratio}
\begin{figure}[ht]
\centering
\includegraphics[width=0.38\textwidth]{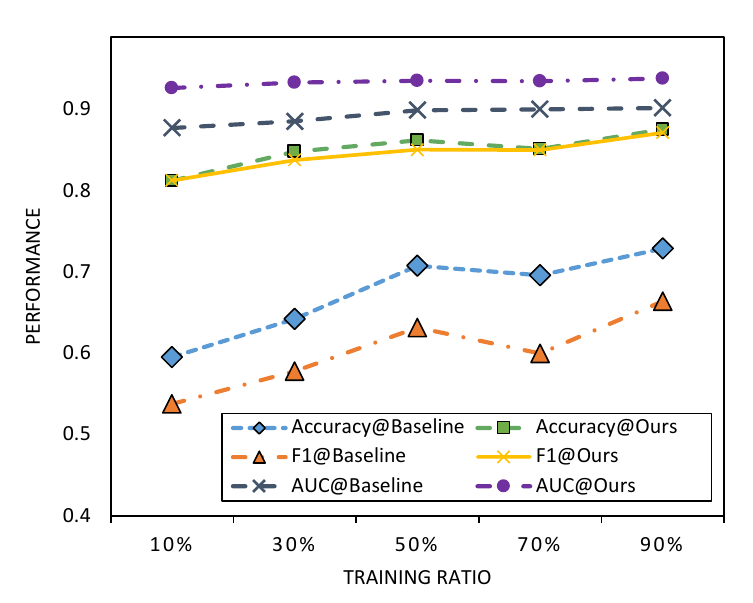}\vspace{-1.3em}
\caption{Effectiveness on Labeled Ratios }
\label{fig: label_ratio}
\end{figure}

To elucidate the impact of training data volume on model proficiency, we conducted experiments with varying proportions of the training set. The outcomes of our methods and the best baseline, delineated in Figure \ref{fig: label_ratio}, underscore the significance of training data volume. A positive correlation emerges between the extent of training data employed and the model's performance metrics. Optimal performance is observed with the complete dataset, indicating that a larger corpus of data typically enhances the model's learning and generalization capabilities.

Nonetheless, our model demonstrates commendable resilience when trained with reduced data, maintaining substantial accuracy and AUC even at 50\% data availability. This resilience is indicative of the model's ability to learn efficiently from a condensed, yet representative, dataset. However, for the baseline, a stark performance decline is noted when the training data is substantially curtailed, as seen in the 10\% training scenario.

\section{related work}
\textbf{Personality Analysis.} The research of social network analysis has been considerably broadened in recent years, incorporating various personality models such as the Five-Factor Model, MBTI\cite{bharadwaj2018persona}, and DISC structure\cite{Lykourentzou2016personalitymatters}. Studies have found personality traits may be associated with intuitive interests \cite{Gao2023hgnn} and health behaviors, such as communication with others \cite{zhang2022correlation}, study-work productivity, and diet \cite{mcnamara2020effect}. In the digital realm, linguistic analysis serves as a foundational tool in these studies, where texts and comments on social media are used to infer personality traits \cite{Li2022languagestyle}, Moreover, multimedia content like videos and photos\cite{sun2022your}, along with interactive metrics (e.g., likes or dislikes), provides deeper insights into user personalities, suggesting a multi-modal approach to personality analysis. Sun et al.\cite{sun2020group} extend this by integrating Network Representation Learning (NRL) to manage multi-class personality classification, demonstrating the model's effectiveness across varied social network distributions. These researches collectively underscore the dynamic interplay of personalities within virtual social environments, paving the way for advanced analytical methodologies. 

\textbf{Graph Neural Networks.}
Graph Neural Networks (GNNs) have been proven effective in dissecting the intricate data structures prevalent in social media. Recent research has proposed and experimented with transformers \cite{Yun2019} using graph or attention networks \cite{yan2020fpgan} for learning graph-structured data, demonstrating their versatility and effectiveness in solving complex problems. Researchers have applied graph network strategies to analyze users' behavior patterns, interactions, and self-reflections \cite{sun2023self}. While GNNs have enabled the modeling of relationships between users, a limitation emerges as each edge connects only limited dimensions, leading to a loss of representation of complex relationships between nodes. As an extension of GNNs, HGNNs (Hypergraph Neural Networks) have emerged as a method that enhances the representation of graph-structured data with higher-level relations \cite{Feng2018, Gao2023hgnn}. Studies have demonstrated the utility of HGNNs in modeling non-pairwise relationships and improving the classification of such data structures within social networks \cite{sun2021heterogeneous}. Recently, researchers have developed a hierarchical two-layer, self-supervised hyperedge neural network to classify users' social networks and a novel graph algorithm to learn and analyze the patterns of users' interactions within social environments \cite{sun2023self}. These advancements highlight the potential of hypergraphs in enhancing the granularity of sociological analysis, offering new perspectives on user interaction patterns and environmental influence.

\section{Conclusion}\label{sec:con}
In this study, we present a sociological personality analysis that integrates hypergraphs with LLMs in online social networks. We offer a very useful dataset to support such thorough analysis. We effectively combine fragmented user data with complex social environments into comprehensive user understanding. Our experimental results demonstrate that this method significantly improves the performance of personality analysis, providing deeper insights into user interactions and their social environments. This work highlights the potential of hypergraph-based models to advance sociological analysis and enrich our understanding of human behavior in digital environments. Future directions include exploring more applications of online personality analysis such as abnormaly detection \cite{sun2022structure}, credit scoring \cite{sun2023counter}, recommendation system \cite{cui2023event}, etc; and exploring other techiniques such as more kinds of relations \cite{zhang2022graph}, graph prompt \cite{sun2023all, sun2023graph}, etc.


\normalem 
\bibliographystyle{unsrt}
\bibliography{cssRef}

\end{document}